\documentclass[aps,prb,twocolumn,showpacs,preprintnumbers,prb,superscriptaddress]{revtex4-1}
\usepackage{graphicx}  
\usepackage{dcolumn}   
\usepackage{bm}        
\usepackage{amssymb}   
\usepackage{framed}
\usepackage{amsmath}
\usepackage{hhline}
\usepackage{color}
\definecolor{bluegreen}{rgb}{0,0.2,0.8}
\usepackage{subfigure,amsmath,verbatim,moreverb}
\usepackage{tabularx}
\usepackage{adjustbox}
\usepackage{lipsum}
\usepackage{longtable}
\usepackage{booktabs}
\usepackage{latexsym}
\usepackage{dcolumn}
\usepackage{amsmath}
\usepackage{epsf}
\usepackage{float}
\usepackage{hyperref}

\usepackage{longtable}
\usepackage{booktabs}
\usepackage{threeparttable}

\usepackage{etoolbox}
\AtBeginEnvironment{align}{\setcounter{subeqn}{0}}
\newcounter{subeqn} %

\usepackage{array}
\newcolumntype{H}{>{\setbox0=\hbox\bgroup}c<{\egroup}@{}}

\newcommand{\Fig}[1]{Fig. \ref{#1}}

\begin{document}
\title{Improving the applicability of the Pauli kinetic energy density based semilocal functional for solids}
\author{Subrata Jana}
\altaffiliation{Corresponding author: subrata.jana@niser.ac.in, subrata.niser@gmail.com}
\affiliation{School of Physical Sciences, National Institute of Science Education and Research, HBNI, 
Bhubaneswar 752050, India}
\author{Sushant Kumar Behera}
\email{sushant@niser.ac.in}
\affiliation{School of Physical Sciences, National Institute of Science Education and Research, HBNI, 
Bhubaneswar 752050, India}
\author{Szymon \'Smiga}
\email{szsmiga@fizyka.umk.pl}
\affiliation{Institute of Physics, Faculty of Physics, Astronomy and Informatics,
Nicolaus Copernicus University, Grudziadzka 5, 87-100 Toru\'n, Poland}
\author{Lucian A. Constantin}
\email{lucian.constantin.68@gmail.com}
\affiliation{Istituto di Nanoscienze, Consiglio Nazionale delle Ricerche CNR-NANO, 41125 Modena, Italy}
\author{Prasanjit Samal}
\email{psamal@niser.ac.in}
\affiliation{School of Physical Sciences, National Institute of Science Education and Research, HBNI, 
Bhubaneswar 752050, India}

\date{\today}

\begin{abstract}
The Pauli kinetic energy enhancement factor $\alpha=(\tau-\tau^W)/\tau^{unif}$ is an important density ingredient, used to construct many meta-generalized gradient approximations (meta-GGA) exchange-correlation (XC) energy functionals, including the very successful strongly constrained and appropriately normed (SCAN) semilocal functional. Another meta-GGA functional, known as MGGAC
[Phys. Rev. B 100, 155140 (2019)], is also proposed in recent time depending only on the $\alpha$ ingredient and based on the generalization of the Becke-Roussel approach with the cuspless hydrogen
exchange hole density. The MGGAC functional is proved to be a very useful and competitive
meta-GGA semilocal functional for electronic structure properties of solids and molecules. Based on the successful implication of the ingredient $\alpha$, which is also useful to construct the one-electron self-interaction free correlation energy functional, here we propose revised correlation energy for MGGAC exchange functional which is more accurate and robust, especially for the high and low-density limits of the uniform density scaling. The present XC functional, named as revised MGGAC (rMGGAC), shows an impressive improvement for the structural and energetic properties of solids compared to its previous version. Moreover, the assessment of the present constructed functional shows to be quite useful in solid-state physics in terms of addressing several current challenging solid-state problems.
\end{abstract}

\maketitle

\section{Introduction}
The Kohn-Sham (KS) density functional theory (DFT)~\cite{kohn1965self} has become 
an indispensable computational framework for performing the electronic structure calculations of new 
generation electronic devices~\cite{geim2013van,novoselov20162d,taylor2001ab}, energy storage 
devices~\cite{lee2016capacitance}, catalyst~\cite{kuklin2019quasiparticle} and spintronics 
devices~\cite{yazyev2008magnetic,lazi2014graphene,cardoso2018van}. Despite the extraordinary potential 
and considerable progress made to date,
challenges remain to bridge the gap between simulations and experimental findings. Efforts have increased 
in developing new strategies in understanding the theory and practical implementation of complex systems, 
but current progress is still far from sufficient. 
However, recent strategies of the development of non-empirical exchange-correlation (XC) functionals by 
satisfying quantum mechanical exact constraints~\cite{sun2015strongly} exhibit extremely 
prosperous to catch the interplay of dimensionality, correlation, charge, orbital character, topological, 
and other large reservoirs of exotic properties of condensed matter systems~\cite{gmitra2015graphene}.

Within the exact quantum constraints, we recall co-ordinate transformations and density scaling rules of 
XC functionals~\cite{levy1985hellmann,
levy2016mathematical,gorling1992requirements,fabiano2013relevance},
second (and fourth) order gradient expansion of XC 
energies~\cite{svendsenPRB96,antoniewiczPRB85,huPRB86,bruecknerPR68,elliottCJC09,elliottPRL08}, low 
and high density limits of the
correlation energy functional~\cite{gorling1994exact,gorling1993correlation,gorling1995hardness}, correct 
asymptotic behavior of the XC energy 
density\cite{MolPhysszs,SMIGA2014125,BUKSZTEL2016263,RSOEP1,DHLUC,constantin2011adiabatic} 
or potential~\cite{dellasalaPRL02,engelZPD92,horowitz2009position,
constantin2016semilocal,niquet2003asymptotic,almbladh1985exact,umrigar1994accurate,
SMIGA2014125,potac1,UevilLUC}, quasi-2D
limit of the XC 
energy~\cite{pollack2000evaluating,kaplan2018collapse,constantin2016simple,constantin2008dimensional}, 
and exact properties of the XC 
hole~\cite{tao2016accurate,tao2008exact,pvrecechtvelovaJCP14,pvrecechtvelovaJCP15}. All these constraints 
are taken into care in different rungs (recognized as the Jacob Ladder~\cite{perdew2001jacob}) of the 
semilocal density functionals approximations (DFA) starting from the local density approximation 
(LDA)~\cite{perdew1992accurate}, throughout generalization gradient approximation 
(GGA)~\cite{perdew1981self,becke1988density,lee1988development,perdew1992atom,
perdew1996generalized,
armiento2005functional,wu2006more,constantin2011improving,zhao2008construction,constantin2015gradient,
fabiano2013testing,constantin2009exchange,cancio2018fitting,
albavera2020generalized} to meta-generalization gradient approximation 
(meta-GGA)~\cite{becke1989exchange,voorhish1998novel,zhao2006new,perdew1999accurate,
tao2003climbing,perdew2009workhorse,constantin2013metagga,
sun2013semilocal,sun2015semilocal,ruzsinszky2012metagga,
constantin2016semilocal,yu2016mn15l,sun2015strongly,tao2016accurate,
wang2017revised,mezei2018simple,jana2019improving,patra2019efficient,patra2019laplacian,
patra2019relevance,smiga2017laplacian,furness2019enhancing,furness2020accurate,
aschebrock2019ultranonlocality}. 
The recent trends in DFT showed that the latter family of 
DFAs can offer unprecedented performance, becoming 
the most auspicious choice of doing both the quantum chemical and material 
simulations~\cite{haas2009calculation,tran2016rungs,metasub1,
metasub2,tao2016accurate,yang2016more,peng2016versatile,mo2017assessment,
tian2017energetic,tian2017accurate,mo2017comparative,
mo2017performance,patra2017properties,peng2017synergy,zhang2017comparative,
sengupta2018from,shani2018accurate,Tang_2018,mo2018accurate,
jana2018assessment,jana2018assessing,patra2019performance,patra2019rethinking,
jana2020accurate,patra2020efficient,mejia2020metagga,
zhang2020localized,tran2020shortcomings,wang2020m06sx}.


We recall that the meta-GGA XC energy functionals have the general form
\begin{equation}
E_{xc}[\rho_{\uparrow},\rho_{\downarrow}]=\int~d{\bf{r}}~\rho({\bf{r}})\epsilon_{xc}^{LDA}
F_{xc}(\rho_{\uparrow},\rho_{\downarrow},\nabla\rho_{\uparrow},\nabla\rho_{\downarrow},
\tau_{\uparrow},\tau_{\downarrow})~,
 \label{introeq1}
\end{equation}
where $\epsilon_{xc}^{LDA}$ is the LDA XC energy per particle, 
$F_{xc}$ is the XC enhancement factor, and 
$\rho_\sigma$ and $\tau_\sigma$
are the $\sigma$-spin density and kinetic energy density, respectively.
Eq. (\ref{introeq1}) is quite flexible, being able to recognize covalent, 
metallic, and weak bonds 
\cite{PhysRevLett.111.106401}. Moreover, the utilization of KS 
kinetic energy density ($\tau({\bf{r}})=\frac{1}{2}\sum_{i,\sigma}|\nabla\psi_{i,\sigma}|^2$, 
$\tau=\tau_\uparrow+\tau_\downarrow$) and different ingredients built in 
from $\tau({\bf{r}})$, density ($\rho=\rho_\uparrow+\rho_\downarrow$), and gradient 
of density ($\nabla\rho$) allow to recover important exact conditions of the XC functional 
\cite{PhysRevLett.91.146401,sun2015strongly,constantin2016semilocal}. Among them, 
we recall 
$\alpha=(\tau-\tau^{W})/\tau^{unif}$, also known as the Pauli kinetic 
energy\cite{levy1988exact,holas1991construction,finzel2018fragment,PauliMeth,constantin2019semilocal} 
enhancement factor ($F_s^\theta)$, is used to construct
several modern XC functionals~\cite{doi:10.1063/1.4789414,PhysRevLett.111.106401,
sun2015semilocal,sun2015strongly,wellendorff2014mbeef}. 
Here $\tau^W=|\nabla \rho|^2/(8\rho)$ and $\tau^{unif}=(3/10)(3\pi^2)^{2/3}\rho^{5/3}$ are the von 
Weizs\"{a}cker and Thomas-Fermi kinetic energy densities, respectively.

The $\alpha$ ingredient helps to impose some exact 
constraints in the functional form, such as the strongly tightened bound of exchange, i.e. $F_x \leq 
1.174$~\cite{lieb1981improved,perdew2014gedanken,sun2015strongly}. Importantly, this allows a  
negative slope ($\partial F_x /\partial \alpha \le 0$) which is related to an improvement in the prediction 
of the band gap~\cite{aschebrock2019ultranonlocality}. Also, the use of solely $\alpha$ and $s=|\nabla 
\rho|/[2(3\pi^2)^{1/3}\rho^{4/3}]$ (reduced density gradient) makes the meta-GGA XC functional free from 
the order-of-limit anomaly, that is important for the structural phase transition pressures and 
energies~\cite{ruzsinszky2012metagga,furness2020examing}.  Focusing on the XC functional development, a popular non-empirical meta-GGA XC functional is the Strongly Constrained and Appropriately
Normed (SCAN)~\cite{sun2015strongly} semilocal DFA.
Besides the SCAN, it was also recently proposed the MGGAC XC functional \cite{patra2019relevance},
whose exchange part depends only on the $\alpha$ ingredient, being developed from a generalization 
of the Becke-Roussel approach \cite{PhysRevA.39.3761,patra2019efficient}, and using the cuspless hydrogen model for the exchange hole density.  
The MGGAC functional showed its productive power over other meta-GGAs 
in terms of the band gap performance for bulk solids~\cite{patra2020electronic}, two dimensional (2D) van 
der Waals (vdW) materials~\cite{geim2013van,novoselov20162d,patra2020electronic}, and structural phase 
stability of solids~\cite{mggac_study}.

The MGGAC exchange enhancement factor is a simple monotonic function of $\alpha$ and satisfies $(i)$ the strongly tightened bound of exchange $F_x 
\leq 1.174$ ($F_x = 1.174$ for two-electron systems), 
$(ii)$ the uniform electron gas limit ($F_x=1$) when $\alpha=1$, and $(iii)$ the cuspless hydrogen related behavior ($F_x=0.937$) at 
$\alpha\to \infty$. On the other hand, the MGGAC correlation energy functional is based on the GGA PBE correlation form \cite{perdewPRL96}, where  
the linear response parameter ($\beta (r_s)=0.030$) was fitted to the equilibrium lattice constants of 
several bulk solids. Then, it is not one-electron self-interaction 
free, an important constraint that should be satisfied by every meta-GGA functional. 
Nevertheless, the MGGAC DFA has demonstrated several successes which include 
$(i)$ improved thermochemical properties~\cite{patra2019relevance}, $(ii)$ improved fundamental band gap 
for bulk and layer solids~\cite{patra2020electronic}, and $(iii)$ correct energy ordering for challenging 
polymorphs of solids~\cite{mggac_study}. 
However, the functional has its own limitations such as moderate performance for lattice constants, bulk 
moduli, and cohesive energies for bulk solids~\cite{patra2019relevance}.
Therefore, it is desirable to improve the accessibility of the MGGAC functional for solids by 
removing some deficiencies of the current version.
In order to do so, 
we keep the MGGAC exchange unchanged, 
modifying only the correlation part of DFA energy such that it becomes exact for one electron 
densities and more robust for low- and high- density limits.

The paper is organized as follows: in the next section, we will construct the revised MGGAC correlation 
energy functional dependent on both $\alpha$ and $s$.
Later, we will couple the proposed correlation with the MGGAC exchange to assess the functional 
performance for molecular and solid-state properties. Furthermore, we also investigate the low- and high- 
density limit of the XC functional. Lastly, we summarize our findings.

\section{Correlation functional construction}
In this section, we propose the revised MGGAC (rMGGAC) correlation energy functional. 
The rMGGAC correlation functional is constructed by incorporating the following important constraints: 
$(i)$ self-interaction free for one-electron systems; $(ii)$ accurate for two electron systems;
$(iii)$ accurate for slowly varying densities; and $(iv)$ accurate spin-dependence in the low-density or strong-interaction limit.
While constraint $(iii)$ is satisfied by the MGGAC correlation energy functional~\cite{patra2019relevance} (i.e., from the PBE form), conditions $(i)$, $(ii)$, and $(iv)$ are not. 
To achieve this goal, we propose the rMGGAC correlation functional, dependent on $\alpha$ 
and $s$ such as, 
\begin{equation}
\epsilon_c^{rMGGAC}=\epsilon_c^{0}f_1(\alpha,s)+\epsilon_c^{1}f_2(\alpha,s)~,
\label{eq1}
\end{equation}
where
\begin{eqnarray}
 \epsilon_c^1&=&\epsilon_c^{LSDA1}+H_1\nonumber\\
 \epsilon_c^0&=&(\epsilon_c^{LDA0}+H_0)G_c(\zeta)~.
 \end{eqnarray}
Here, $\epsilon_c^{LSDA1}$ is the Perdew-Wang 1992 (PW92) LSDA correlation~\cite{perdew1992accurate}, and $\epsilon_c^{LDA0}G_c(\zeta)$ is the 
local correlation from SCAN functional~\cite{sun2015strongly}, that has been shown to be accurate for 
two electron systems~\cite{sun2015strongly,sun2016communication}. 
Here $\zeta=(\rho_\uparrow-\rho_\downarrow)/\rho$ is the relative spin polarization. 
Note that it does not contain the gradient of the density, having the following 
expressions~\cite{sun2015strongly}
\begin{equation}
\epsilon_c^{LDA0}=-b_{1c}/(1+b_{2c}r_s^{1/2}+b_{3c}r_s),
\label{eq2ll}
\end{equation}
and 
\begin{eqnarray}
& G_c(\zeta)=\{1-2.3631 [d_x(\zeta)-1]\}(1-\zeta^{12}),\\
& d_x(\zeta)=[(1+\zeta)^{4/3}+(1-\zeta)^{4/3}]/2~.
\label{eq2}
\end{eqnarray}
Here, we utilize $H_0=b_{1c}\ln[1+w_0(1-g_{\infty}(\zeta=0,s))]$ and 
$H_1=\gamma\phi^3\ln[1+w_1(1-g(At^2))]$~\cite{sun2015strongly},
with 
$w_0=\exp[-\epsilon_c^{LDA0}/b_{1c}]-1$, $g_{\infty}=1/(1+4\chi_{\infty}(\zeta=0)s^2)^{1/4}$, 
$w_1=\exp[-\epsilon_c^{LDA1}/(\gamma\phi^3)]-1$,
$g(At^2)=1/(1+At^2)^{1/4}$, $A=\beta/(\gamma w_1)$, and $t=(3\pi^2/16)^{1/3}s/(\phi r_s^{1/2})$  (a 
reduced density gradient for 
correlation) from SCAN correlation~\cite{sun2015strongly}. 
The used terms and parameters are as follows: $b_{1c} = 0.0285764$, $b_{2c} = 0.0889$, $b_{3c} = 0.125541$,
$g_{\infty}(\zeta=0,s)=1/(1+4\chi_\infty(\zeta=0) s^2)^{1/4}$, $\chi_\infty(\zeta=0)=0.128026$, 
$\gamma = 0.031091$, and $\phi=[(1+\zeta)^{2/3}+(1-\zeta)^{2/3}]/2$. 
The readers are suggested to go through the supporting information of ref.~\cite{sun2015strongly} for 
details of the terms and parameters values constructed. 
We use $\beta=0.066725$, which is the correct second-order coefficient in the high-density limit, 
also present in PBE 
correlation~\cite{perdew1996generalized}.

Finally, we construct the functions $f_1(\alpha,s)$ and $f_2(\alpha,s)$, from the following conditions:

$(i)$ For the slowly varying density limit, which is recognized as $\alpha\approx 1$ and $s\approx\ 0$,
$\epsilon_c^{rMGGAC}$ becomes $\epsilon_c^1$.

$(ii)$ For one electron or two electron singlet state, which is recognized as $\alpha\approx 0$, $\epsilon_c^{rMGGAC}$ becomes $\epsilon_c^0$.

To satisfy these two conditions, we propose following ansatz
\begin{eqnarray}
& f_2(\alpha,s)=\frac{3[g(\alpha,s)]^3}{1+[g(\alpha,s)]^3+[g(\alpha,s)]^6}\\
& f_1(\alpha,s)=1-f_2~,\\
\label{eq3}
\end{eqnarray}
where
\begin{equation}
 g(\alpha,s)=\frac{(1+\gamma_1)\alpha}{\gamma_1+\alpha+\gamma_2 s^2}~.
\end{equation}
The parameters $\gamma_1=0.08$ and $\gamma_2=0.3$ are chosen from the atomization energies of the AE6 molecules~\cite{haunschild2012theoretical}.
Note that in the tail of the density, if the valence atomic orbital is degenerate, as in the case of Ne atom, then $\alpha\to\infty$, 
else $\alpha\to 0$ as in case of alkali and alkaline-earth atoms.

\begin{figure}
\includegraphics[width=8 cm, height=6 cm]{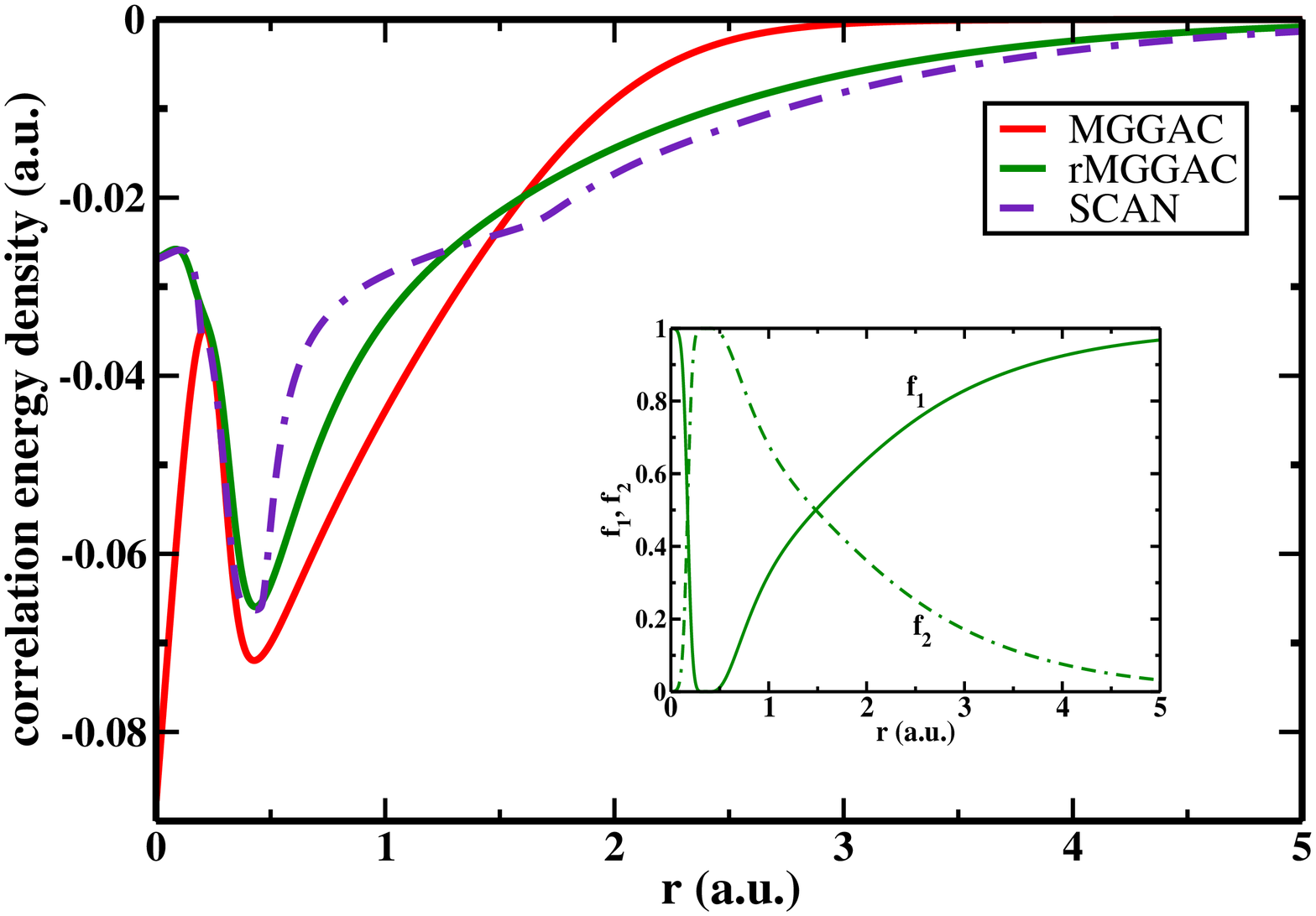}
\caption{Correlation energy per particle $\epsilon_c$ versus the radial distance $r$, for Ne atom. In the 
inset we show the functions $f_1$ and $f_2$.
}
\label{figec}
\end{figure}
In Fig. \ref{figec} we show a comparison between rMGGAC, MGGAC and SCAN correlation energies per particle, for Ne atom. By construction, rMGGAC recovers $\epsilon_c^0$ at the nucleus (where $\alpha\approx 0$) and in the tail of the density (where $s$ diverges), while it recovers $\epsilon_c^1$ in the atomic core (where the density is compact and slowly-varying). Overall, rMGGAC agrees closely with, but is smoother than SCAN, being significantly different from the MGGAC behavior.   

We mention that the choice of the $f_1(\alpha,s)$, $f_2(\alpha,s)$, and $\epsilon_c^{rMGGAC}$ keeps the correct constraint of the 
correlation energy functional. Such as, in the rapidly varying density limit, recognized as $t\to \infty$ and $s\to\infty$, 
both $H_0$ and $H_1$ correctly cancel with $\epsilon_c^{LSDA0}$ 
and $\epsilon_c^{LSDA1}$, respectively, making the correlation vanishes, an important constraint satisfied 
by all functionals constructed on the top of PBE form~\cite{perdew1996generalized}.
Nevertheless, as shown in Fig. \ref{figec}, this vanishing process is much slower for rMGGAC and SCAN than for MGGAC.
Here we also note that in the tail of the density, if the valence atomic orbital is degenerate, as in the case of Ne atom, then $\alpha\to\infty$, 
else $\alpha\to 0$ as in case of alkali and alkaline-earth atoms.

\begin{figure}
   \includegraphics[width=6.0 cm]{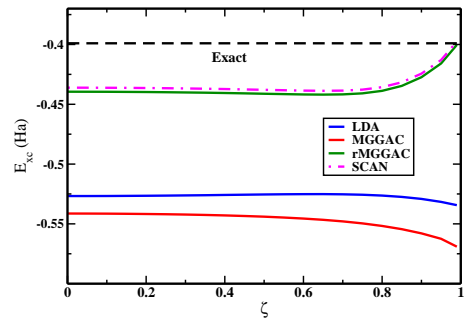}
  \caption{Exchange-correlation energy in the low-density
limit for spin-polarized one electron Gaussian 
density.}
  \label{gaussina}
\end{figure}
In the low-density limit (see APPENDIX A for details), the rMGGAC XC energy functional becomes independent of $\zeta$ for $0\leq\zeta\leq 0.7$ 
(as shown in Fig.~\ref{gaussina} for one electron Gaussian density). 
The $\zeta$ independence of the XC functional is an important constraint not only for Wigner crystals, but also for the atomization energies~\cite{perdew2004meta}.

\begin{table}
\begin{center}
\caption{\label{corr} Error statistics (mean absolute error (MAE) and mean absolute relative error (MARE)) 
of the correlation energy per electron ($E_c/N$ in mHa) for 25 atoms and ions. Full results are reported in Ref. \cite{jana2020supplimentary}.
}
\begin{tabular}{cccccccc}
\hline\hline
Error& SCAN  &  MGGAC  &  rMGGAC  \\
\hline
MAE&3.7&10.5&3.4\\
MARE$^a$&11.3&48.5&11.0\\
\hline\hline
\end{tabular}
\begin{tablenotes}
            \item[]a) computed without the H atom.
\end{tablenotes}
\end{center}
\end{table}
To quantify the performance of correlation functional for high-density limit, in Table~\ref{corr}, we consider the correlation energies 
for the benchmark test of several atoms and ions, also used in Ref. \cite{constantin2019correlation}. It is shown that the rMGGAC 
correlation energy considerably improves over MGGAC, being in line with the SCAN correlation.

\begin{figure*}
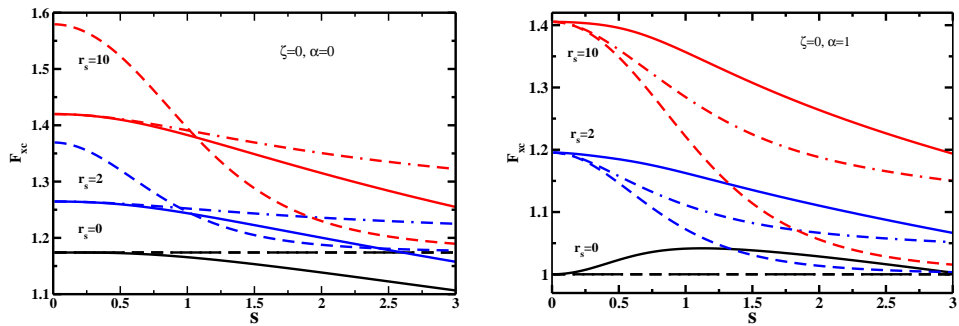

\includegraphics[width=6 cm]{Fxc_alpha_0.eps}
\hspace{0.5 cm}
\includegraphics[width=6 cm]{Fxc_alpha_1.eps}
\caption{Shown is the $F_{xc}$ versus the reduced gradient $s$, for SCAN (solid), MGGAC (dashed), and 
rMGGAC (dashed-dotted) functionals,
considering $\zeta=0, \alpha=0$ (upper panel) and $\zeta=0, \alpha=1$ (lower panel), and several values 
of the bulk parameter ($r_s=0,\; 2$, and $10$).
}
\label{fighex1}
\end{figure*}
\begin{figure}[ht]
\includegraphics[width= 6 cm]{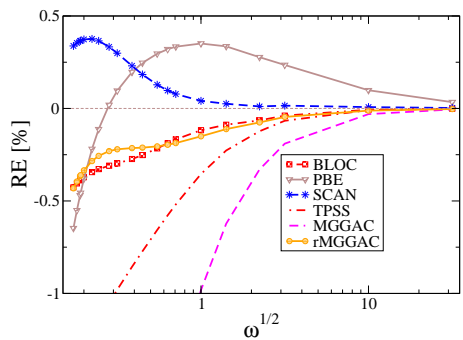}
\caption{Relative error on total energies of harmonium atoms for various values of confinement strength $\omega$.}
\label{fig2}
\end{figure}
Finally in Fig. \ref{fighex1} we compare the SCAN, MGGAC, and rMGGAC XC enhancement factors $F_{xc}(\zeta,\alpha,s,r_s)$, for the spin-unpolarized choice ($\zeta=0$) and $\alpha=0$ (that is describing the two-electron singlet states) and $\alpha=1$ (the case of slowly-varying density limit when $s\le 1$). In both panels, the $r_s=0$ represents the exchange-only (high-density) limit, and we observe that MGGAC (rMGGAC) enhancement factor is independent on $s$ ($F_x^{MGGAC}=1.174$ for $\alpha=0$, and $F_x^{MGGAC}=1$ for $\alpha=1$). Overall, the rMGGAC XC enhancement factor is slightly bigger than SCAN for $\alpha=0$, and slightly smaller than SCAN $\alpha=1$. We also note the large differences between MGGAC and rMGGAC, especially for the low-density cases ($r_s=10$).    


\begin{table*}
\caption{\label{solids} The test set of 41 solids
considered in this work for lattice constants, bulk moduli, and cohesive energies. The space group is indicated in parenthesis.}
\begin{ruledtabular}
\begin{tabular}{ccccccccccccccccc}
~~~~~~~~~~~~~~~~Solids for lattice constants, bulk moduli, and cohesive energies \\
\hline
Li ($Im\overline{3}m$), Na ($Im\overline{3}m$), Ca ($Fm\overline{3}m$), Sr ($Fm\overline{3}m$), Ba ($Im\overline{3}m$), Al ($Fm\overline{3}m$),
Cu ($Fm\overline{3}m$), Rh ($Fm\overline{3}m$), Pd ($Fm\overline{3}m$), \\
Ag ($Fm\overline{3}m$), Au ($Fm\overline{3}m$), Cs ($Im\overline{3}m$),
Fe ($Im\overline{3}m$), Ir ($Fm\overline{3}m$), K ($Im\overline{3}m$), Mo ($Im\overline{3}m$), Nb ($Im\overline{3}m$), Ni ($Fm\overline{3}m$), \\
Pt ($Fm\overline{3}m$), Rb ($Im\overline{3}m$), Ta ($Im\overline{3}m$), V ($Im\overline{3}m$), W ($Im\overline{3}m$), C ($Fd\overline{3}m$), 
Si ($Fd\overline{3}m$), Ge ($Fd\overline{3}m$), SiC ($F\overline{4}3m$), \\
GaAs ($F\overline{4}3m$), AlAs ($F\overline{4}3m$), AlN ($F\overline{4}3m$), AlP ($F\overline{4}3m$),
GaN ($F\overline{4}3m$), GaP($F\overline{4}3m$), InAs ($F\overline{4}3m$), InP ($F\overline{4}3m$), \\
InSb ($F\overline{4}3m$), LiF ($Fm\overline{3}m$),
LiCl ($Fm\overline{3}m$), NaF ($Fm\overline{3}m$), NaCl ($Fm\overline{3}m$), MgO ($Fm\overline{3}m$)
\end{tabular}
\end{ruledtabular}
\end{table*} 


\section{Results and Discussions}
In this section, we present the assessment of the performance of the rMGGAC XC DFA,
along with SCAN and MGGAC methods for harmonium atom, 
various molecular and solid-state problems including the 
band gap assessment and structural phase transition of challenging systems.

\begin{table}
\begin{center}
\caption{\label{tabab1} The error statistics (MAEs and MAREs) of equilibrium lattice constants (LC) 
($a_{0}$ in {m\AA}), bulk 
modulus (BM) ($B_{0}$ in GPa), and cohesive energies (COH) ($\epsilon_{\rm{coh}}$ in eV/atom) of 41 
solids. 
For band gaps ($\epsilon_g$) we consider 40 bulk semiconductors which includes the $31$ semiconducting systems band gaps from
SBG31 test of Ref.~\cite{verma2017hle16} and additionally we include the band gaps of $9$ solids (Cu$_2$O, CuBr, ScN, SnTe, MgO, 
NaCl, LiCl, NaF, and LiF). The MARE of the SBG31 test is also reported. For Surface energies 
($\epsilon_{surface}$ (in J/m$^2$)) 
we consider the (111) surface. We also calculate the CO molecule adsorption energies ($\epsilon_{ads}$)  
on top of 
the (111) surface (in eV). For both cases we consider five different 
transition metals taken from Ref.~\cite{patra2019relevance}. The SCAN values are taken from 
Ref.~\cite{mejia2020metagga}. The surface energies and adsorption energies of SCAN and MGGAC functionals are taken from ref.~\cite{patra2019relevance}. 
The LC20, BM20, COH20, and SBG31 test sets results of MGGAC 
are taken from ref.~\cite{patra2019relevance}. See ref.~\cite{jana2020supplimentary} for the details of 
the 
rMGGAC values.}
\begin{ruledtabular}
\begin{tabular}{cccccccccccccccc}
                  & SCAN & MGGAC & rMGGAC  \\
\hline
\multicolumn{4}{c}{Lattice constants} \\
MAE (m\AA)&38&51&38        &  \\
MARE (\%)&0.81&1.09&0.81 \\
MAE of LC20 (m\AA)     &25&45&30         &  \\
\multicolumn{4}{c}{Bulk moduli} \\
MAE (GPa)  &7.5&11.5&8.5  \\
MARE (\%) &6.3&11.2&8.8\\
MAE of BM20 (GPa)     &4.2&10.0&4.6         &  \\
\multicolumn{4}{c}{Cohesive energies} \\
MAE (eV/atom)  &0.19&0.36&0.30  \\
MARE (\%) &4.80&9.29&8.11\\
MAE of COH20 (eV/atom)     &0.11&0.38&0.27  \\
\multicolumn{4}{c}{Band gaps} \\
MAE (eV)&1.1&0.8&0.5  \\
MARE (\%)&39.2&40.6&43.3 \\
SBG31  &38.1&40.7 &42.6  \\
\multicolumn{4}{c}{Surface energies} \\
MAE (eV)  &0.48&0.21&0.27  \\
MARE (\%) &22.8&10.5&12.2\\
\multicolumn{4}{c}{Adsorption energies} \\
MAE (J/m$^2$) &0.51&0.28&0.31  \\
MARE (\%)&42.1&23.7&24.3 \\
\end{tabular}
\end{ruledtabular}
\end{center}
\end{table}

\subsection{Model system: Harmonium atom}

We have tested the performance of the several DFAs (including the rMGGAC) against the harmonium atom\cite{PhysRev.128.2687}, for various 
values of the confinement strength $\omega$. 
We recall, that at small values of $\omega$, the system is strongly correlated, whereas for large values of $\omega$ it is tightly bounded. These two regimes are very important for many condensed matter applications. Hence, the harmonium atom provides an excellent tool for testing approximate density functional methods\cite{Test1,Test3}. Similarly to our past studies \cite{LUCISI,SRhSG4}, in order to perform calculations we have utilized an even-tempered Gaussian basis set from Ref. \onlinecite{B926389F} for $\omega \in [0.03, 1000]$. As a reference results, we have used full configuration interaction (FCI) 
data  obtained in the same basis set which have been proved to be close to exact values\cite{LUCISI}.%

In \Fig{fig2}, we report the relative error (RE) on total energies calculated with respect to reference data for several XC functionals as a function of $\omega$. All DFAs energies have been obtained in post-SCF fashion on top of self-interaction free OEPx (optimized effective potential exact exchange) densities as in Refs. \onlinecite{LUCISI,SRhSG4}. One can note the remarkable good performance of rMGGAC functional along with the whole range of $\omega$  with an overall MARE of 0.21\%. Moreover, rMGGAC significantly outperforms the MGGAC (MARE=2.67\%), being especially visible for the strongly correlated regime. This may indicate that the new correlation functional form is more compatible with MGGAC exchange. The rMGGAC behavior is also quite similar to BLOC functional\cite{BLOC} which also underestimate slightly the reference results with an overall MARE of 0.23\%. The SCAN DFA gives similar results (MARE=0.20\%), although in this case, we observe the overestimation. For comparison, we also show the TPSS and PBE data which gives an overall MARE of 0.78\% and 0.28\%, respectively.

\subsection{General assessment for solids}

%
%
%
\begin{figure}
  \includegraphics[width= 12 cm, angle =270]{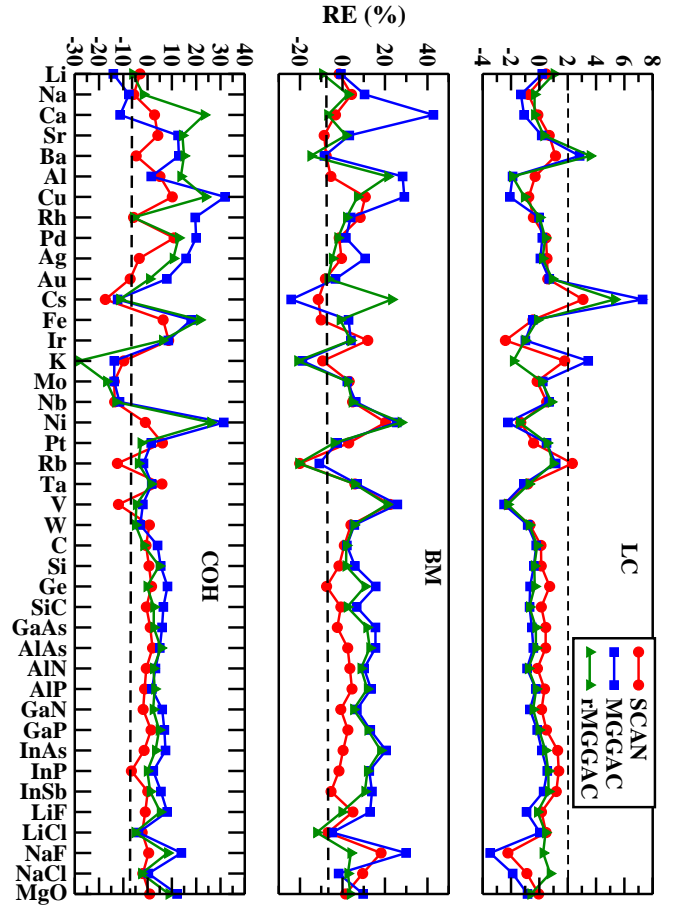}
  \caption{Shown is the RE (\%) of LC, BM, and COH test sets as obtained from 
  different methods.}
  \label{plotall}
\end{figure}

For general assessment,
we first consider the $41$ bulk solids set proposed in Table~\ref{solids} 
and we compute the equilibrium lattice constants (LC41), bulk moduli (BM41), and cohesive energies 
(COH41). This benchmark test set was also used to assess the
performance of SCAN~\cite{sun2015strongly} and MGGAC~\cite{patra2019relevance}
functionals.
We report the error statistics (MAE and MARE)
for all the solids in TABLE~\ref{tabab1}.  Our benchmark calculations for the LC41 test  indicate that 
the rMGGAC functional 
gives the MAE $\approx 38$ m\AA, which is a remarkable improvement performance compared to 
MGGAC, being close to the state-of-art accuracy of the SCAN meta-GGA.
We also report the MAE of the LC20 test set compiled in ref.~\cite{sun2011self} for which MAEs for other methods (not included here) are also 
available. From literature we found other meta-GGAs, like the revised Tao-Perdew-Staroverov-Scuseria 
(revTPSS)~\cite{perdew2009workhorse} and Tao-Mo (TM)~\cite{tao2016accurate}
semilocal functionals, that give for LC20 test, the MAE $\approx 32$ 
m\AA~\cite{sun2011self,jana2019improving}.

The improvement of the lattice constants from rMGGAC functional is also followed by its performance for 
the bulk moduli, showing that the rMGGAC predicts the realistic shape of the energy versus volume curve near 
the equilibrium point. Thus, we obtain for the BM41 test, an 
improved MAE $\approx 8.5$ GPa
from rMGGAC compared to the MGGAC (MAE $\approx 11.5$). Our benchmark values are also close to that 
of the SCAN where the 
later gives MAE of $7.5$ GPa. For BM20 test set we obtain MAE of 4.6 GPa from rMGGAC. The performance of the 
rMGGAC for BM20 is also better or comparable than the state-of-art GGA for solids, PBEsol (MAE 6.2 
GPa)~\cite{sun2011self} and meta-GGA revTPSS (MAE 8.7 GPa)~\cite{sun2011self}
and TM (MAE 4.2 GPa)~\cite{jana2019improving}.

The results for the COH41 test also display the improved performance from rMGGAC (MAE of $0.30$ eV/atom) 
over the MGGAC (MAE of $0.36$ eV/atom). Here, the SCAN (MAE of $0.19$ eV/atom) performs better 
than rMGGAC. However, our reported MAE values for COH20 indicates that rMGGAC values are close to the GGA PBEsol (MAE 0.25 eV/atom)
~\cite{sun2011self} and TM meta-GGA semilocal functionals (MAE 0.28 eV/atom)~\cite{jana2019improving}. 
Note that for cohesive energies,
well balanced performances for solids and atoms are required.

The different functional performances for the general purpose bulk solid properties are 
also shown in 
Fig.~\ref{plotall}, where we observe an almost systematic improvement of rMGGAC over the MGGAC.
For lattice constants, the Cs is challenging, for which we observe improvement from MGGAC to rMGGAC and 
to SCAN.
Note that being a soft matter (i.e. small bulk moduli systems), the accurate description of the lattice 
constants 
of Cs needs a vdW interaction. Here, all methods overestimate the lattice constants indicating the lack 
of vdW
interaction in the functional form. For cohesive energies and bulk moduli, we observe the magnetic Fe 
and Ni systems 
are bit of challenging for rMGGAC. Note that all these functionals overestimate the magnetic moments of 
the itinerant electron 
ferromagnets~\cite{mejia2018deorbitalized,mejia2019analysis,mejia2020metagga,
fu2018applicability,tran2020shortcomings} which is probably 
the source of error for bulk moduli and cohesive energies of Fe and Ni. Here, we observe that SCAN is 
bit better than MGGAC and rMGGAC for Ni cohesive energy.

Next, we consider the band gaps of $40$ bulk semiconductors which is constructed from 
the $31$ semiconducting band gaps of 
SBG31 test taken from Ref.~\cite{verma2017hle16}, additionally we include the band gaps of $9$ solids: 
Cu$_2$O, CuBr, ScN, SnTe, MgO, 
NaCl, LiCl, NaF, and LiF. All band gaps are calculated at the experimental lattice constants, where the 
lattice parameters are taken from 
ref.~\cite{tran2017important}. The rMGGAC performance on band gaps of bulk solids is quite close 
to that of the MGGAC. Our test set consists of narrow to wide band gap 
insulators.

\begin{figure}
  \includegraphics[width= 8 cm]{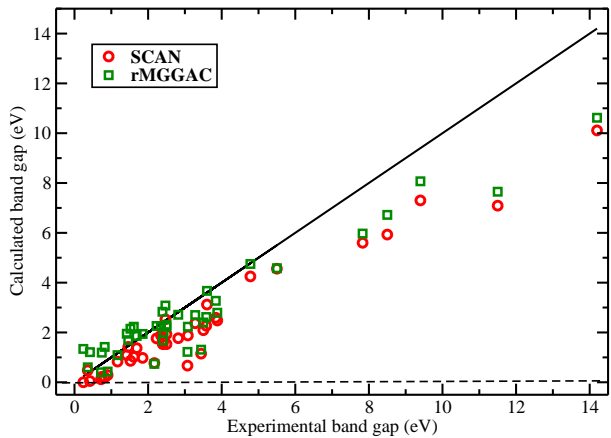}
  \caption{Shown is the calculated versus experimental band gap for SCAN and rMGGAC 
functionals for 40 bulk solids considered in 
  this work. We do not show MGGAC as its band gaps are practically same as rMGGAC.}
  \label{band-plot}
\end{figure}

Interestingly, we observe that MGGAC and rMGGAC give non-zero band gap even for 
InSb, a difficult case for SCAN, where 
the generalized KS (gKS) gap of SCAN is zero. 
We obtain 1.34 eV gap for InSb from rMGGAC which is larger 
than the experimental reported value (0.24 eV). The readers are suggested to go through 
refs.~\cite{tran2019semilocal,borlido2019large,Borlido2020xc} for other different functionals 
performance in bandgap assessments. Nevertheless,
the improvement of the MGGAC and rMGGAC band gaps over SCAN method for band gap of solids within $1.5<E_g<6$ eV 
are clearly visualized from Fig.~\ref{band-plot}. The improvement 
in the performance of MGGAC and rMGGAC is due to the use of only $\alpha$ ingredient in the exchange
functional form, which gives more negative slope $\partial F_x/\partial\alpha$, achieving a stronger 
ultra-non-locality 
effect \cite{aschebrock2019ultranonlocality}.
However, we mention that both MGGAC and rMGGAC overestimate the band gap of low-gap semiconductors, 
having MARE worse than SCAN.

Finally, we consider the  surface energies and on top CO surface adsorption energies for (111) surfaces 
of transition metals Cu, Pd, Pt, Rh, and Ir. 
These tests are 
particularly interesting because they involve transition metal surface 
phenomena, that are important for catalytic applications. We observe that without including any 
van-der Waals (vdW) long-range corrections, MGGAC and rMGGAC perform slightly
better than SCAN. We recall that SCAN may perform better with a vdW 
non-local correction~\cite{patra2017properties,sharada2019adsorption}.

\begin{table}
\scriptsize
\begin{center}
\caption{\label{ta3} Band gap of different solid structures (TMDC in their bulk construction and monolayers) as obtain from SCAN,
MGGAC, and rMGGAC methods. For TMDC bulk the PBE spin orbit coupling (SOC) corrections (taken from 
Ref.~\cite{patra2019efficient}) are subtracted from each methods. 
We also consider SOC corrections for monolayers. 
}
\begin{tabular}{cccccccc}
\hline\hline
 Solids         &Exp.   & SCAN  &  MGGAC  &  rMGGAC  \\
 \hline
\multicolumn{6}{c}{Bulk TMDC$^a$}\\[0.2 cm]
HfS$_2$	        &1.96	&1.29   & 1.51        &	1.50\\
HfSe$_2$	&1.13	&0.68   & 0.92        &	0.90\\
MoS$_2$	        &1.29	&1.01   & 1.08        &	1.06\\
MoSe$_2$	&1.1	&0.96   & 1.03        &	1.02\\
WS$_2$	        &1.35	&1.18   & 1.16        & 1.14\\
WSe$_2$	        &1.2	&1.11   & 1.09        & 1.08\\
ZrS$_2$	        &1.68	&1.12   & 1.38        & 1.36\\
ZrSe$_2$	&1.2	&0.51   & 0.80        & 0.76\\

\multicolumn{6}{c}{Monolayers$^b$}\\[0.2 cm]
h-BN & 5.80~\cite{okada2000border}, 6.00~\cite{paleari2018excitons}&4.80  &5.39  & 5.33 &  & \\
MoS$_2$ & 1.86~\cite{eknapakul2014electronic}, 1.80~\cite{jena2007enhancement}, 2.16~\cite{ugeda2014giant} &1.83  &1.78  & 1.80 &  & \\
MoSe$_2$&1.54~\cite{choi2017temperature}, 1.58~\cite{zhang2014direct}, 1.55~\cite{tongay2012thermally}&1.53&1.48&1.49&\\
MoTe$_2$&1.10~\cite{ruppert2014optical,lezama2015indirect}, 1.02~\cite{lezama2014surface}&1.12 &1.08&1.09&\\
WS$_2$&1.58~\cite{amin2014strain}, 1.57~\cite{zhang2019carbon} &1.71&1.85&1.87&\\
WSe$_2$&1.66~\cite{hsu2017evidence,blundo2020evidence}&1.41&1.51&1.53&\\
WTe$_2$&1.44~\cite{dixit2018structural}&0.73&0.79&0.81&\\
BP & 1.12~\cite{huisheng2020high} &1.19  & 1.53 & 1.39&  & \\


\hline\hline
\end{tabular}
\begin{tablenotes}
            \item[]a) Simulations are performed at  RPA calculated lattice constants~\cite{bjorkman2012vdw,bjorkman2012testing}. For reference setup 
and experimental values of band gaps see TABLE III of 
             Ref.~\cite{patra2019efficient} and all references therein. \\
            \item[]b) Calculated at PBE optimized lattice constants.\\
            \end{tablenotes}
\end{center}
\end{table}

\subsection{Electronic band gap of 2D vdW Transition Metal Dichalcogenide and Monolayers}

In a generalized way, 2D vdW materials offer great flexibility in terms of tuning their electronic 
properties ~\cite{choi2012binary,wang2017porous}. Thus, electronic band-gap engineering can be carried 
out by changing the number of 2D layers in a given material 
~\cite{wu2012three,zhao2014carbon,du2016insilico}. 
The stacking of two or more monolayers of different vdW materials leads to the rich variety of 2D vdW quantum systems and allows 
researchers to explore novel and collective electronic, topological, and magnetic phenomena at the 
interfaces~\cite{martins2007electronic}. The correlation between electron and spin, phonons, and other 
electrons can be significantly modulated affecting the charge transport, entropy, and total energy in the 
quantum limit. 
Especially, the bandgap engineering of 2D vdW systems from a low cost semilocal 
methods is still very difficult to achieve where various methods based on the hybrid density functionals and 
Green functional (GW) level theory is proposed. Also, the GGA functionals perform moderately to estimate 
the bandgaps of these systems, whereas meta-GGA methods are not vastly assessed. In the present context, 
we show a systematically improvable performance for bandgaps of those systems from SCAN to MGGAC and/or rMGGAC 
functionals. To depict this we consider bandgaps of several bulk 2D vdW transition metal dichalcogenide 
(TMDC) and monolayers, which are presented in Table ~\ref{ta3}. These systems 
have electronic band gap ranging from the semiconducting monolayer 2H-MoS$_2$ (1.8 
eV) and black phosphorene (BP) (1.2 eV) to their bulk systems 2H-MoS$_2$ (1.29 eV) to wide gap 
insulator of hexagonal boron-nitride (h-BN) (5.4 eV). The results of the aforementioned systems are presented in TABLE~\ref{ta3}. Inspecting TABLE~\ref{ta3} we 
observe, MGGAC and rMGGAC can be considered as improved methods to address the bandgaps of the bulk TMDC 
to monolayer. 
In all the cases, the 
MGGAC and rMGGAC perform in a quite similar way, predicting the bandgaps better than the SCAN functional. 

\begin{figure}
\includegraphics[width=\columnwidth]{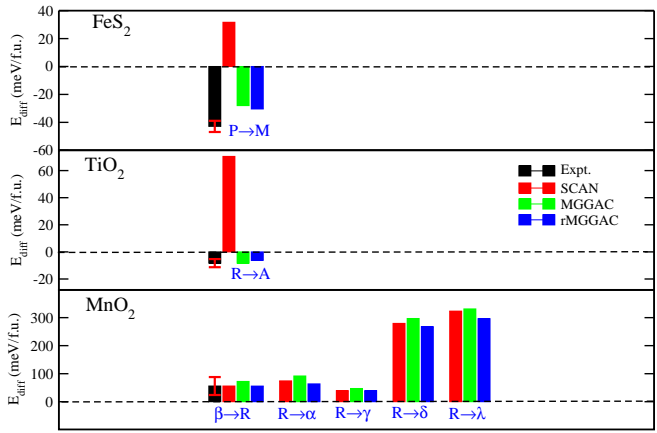}
\caption{Shown is the energy differences of different phases of FeS$_2$, TiO$_2$, and MnO$_2$ polymorphs. 
We do not consider zero-point energy (ZPE) corrections. 
For MnO$_2$, the SCAN values are taken from reference~\cite{kitchaev2016energetics} and 
MGGAC values from reference~\cite{mggac_study}. For experimental values we consider enthalpy differences.
Full results are also reported in Ref. \cite{jana2020supplimentary}.
}
\label{figphase}
\end{figure}
%

\subsection{Transition pressure and structural stability}

\begingroup
\begin{table}
\scriptsize
\caption{Tabulated are the phase transition pressure ($P_t$) (in GPa) of highly symmetric phases. 
The SCAN temperature corrections are added to the MGGAC and rMGGAC ground-state results.
The temperature uncorrected values for all functionals are reported in Ref. \cite{jana2020supplimentary}.
}
\begin{ruledtabular}
\begin{tabular}{lcccccccccccccccccccccccccccccccccc}

Solids	& Expt.$^a$&SCAN$^a$	&MGGAC		&rMGGAC		\\ \hline
Si	&12.0	&13.8		&20.2	&	14.8       	\\
Ge	&10.6	&10.4		&15.8	&	13.7       	\\
SiC 	&100.0	&69.1		&69.2	&	61.6       	\\
GaAs	&15.0	&16.1		&24.1	&	22.9       	\\
Pb    	&14.0	&22.2		&13.7	&	13.1      	\\
C     	&3.7	&8.3		&7.7	&	6.5       	\\
BN  	&5.0	&6.1		&4.4	&	3.7       	\\
SiO$_2$ &7.5    &5.2            &2.0    &       2.6              
\label{tabpt}
\end{tabular}
\begin{tablenotes}
            \item[]a) See Ref.~\cite{sengupta2018from} and all references therein.
            \end{tablenotes}
\end{ruledtabular}
\end{table}
\endgroup

Accurate prediction of the phase transition pressure and correct ground state phase of polymorphs of 
solids are quite important from the point of view of the stability of the solid. 
Importantly, meta-GGA functionals that suffer from the order-of-limit problem, wrongly predict the phase 
transition pressure and structural phase transition of different polymorphs of 
phases~\cite{ruzsinszky2012metagga,patra2020way}. But functionals like SCAN and MGGAC methods do not 
suffer from the order-of-limit problem. In Table~\ref{tabpt}, we report the phase 
transition pressure of different phases of the solids. We notice that the MGGAC has the tendency of 
overestimation 
the phase transition pressure for Si and Ge, while rMGGAC corrects this 
limitation of the MGGAC. This is because the energy differences of different phases are 
better described by rMGGAC than MGGAC functional. Overall, SCAN functional performs better here for 
all transition pressures.

The improved performance of rMGGAC is also noticeable from the structural phase stability of different 
polymorphs of FeS$_2$, TiO$_2$, and MnO$_2$. These systems are known to
be challenging systems for semilocal functionals as those often fail to predict the correct energy 
ordering of these systems~\cite{zhang2018relative,cui2016first,kitchaev2016energetics,mggac_study}. 
While the SCAN functional correctly predicts all phases of MnO$_2$ polymorphs, it wrongly estimates the 
ground states of both FeS$_2$ and TiO$_2$. The SCAN functional predicts marcasite (M) and 
anatase (A) as the most stable phases over the pyrite (P) and rutile (R) phases for FeS$_2$ and TiO$_2$, 
respectively~\cite{zhang2018relative,cui2016first,mggac_study}. 
However, it is shown~\cite{mggac_study}, correct energy ordering of these systems can be achieved from 
the MGGAC level theory. Moreover, we observe from Fig.~\ref{figphase}
that rMGGAC further improves the performance of MGGAC in terms of the energy differences. For MnO$_2$ the 
energy differences as obtained from rMGGAC are very close or even better than those obtained using SCAN functional. 

\subsection{General thermochemical assessment}


\begin{table}
\caption{\label{tabthermo} 
MAEs (in kcal/mol) for Main group thermochemistry (MGT), Barrier heights (BH), and 
Non-covalent interactions (NCI) test sets. The SCAN and MGGAC values from 
reference~\cite{patra2019relevance}. The SCAN and MGGAC values are taken from ref.~\cite{patra2019relevance}.}
{\scriptsize
\begin{tabular}{ccccccccc}
\hline\hline
              &  SCAN &MGGAC &rMGGAC \\
\hline
\multicolumn{4}{c}{Main group thermochemistry (MGT)} \\[0.1 cm]
	AE6$^a$        &  3.43     &5.24   &4.64      \\
	G2/148$^b$      &  3.73    &4.38 &4.61      \\
        EA13$^c$&3.22     &3.16& 4.14     \\
        IP13$^d$&4.43&8.04&5.73\\
        PA8$^e$&1.41&5.59&2.74\\
	DC9/12$^f$  & 11.13   &6.35   &11.75     \\
	HC7$^g$     & 6.51    & 4.10  & 6.63        \\
	BH76RC$^h$       & 2.32           &2.51    &2.63           \\[0.1 cm]
	\multicolumn{4}{c}{Barrier heights (BH)} \\[0.1 cm]	
        HTBH38$^i$  &  7.31     &2.86 &5.13      \\
        NHTBH38$^j$ &  7.88  & 4.95    & 5.29       \\[0.1 cm]
	\multicolumn{4}{c}{Non-covalent interactions (NCI)} \\[0.1 cm]	
        HB6$^k$      & 0.76  &0.87  & 0.79    \\
	DI6$^l$      & 0.53  & 0.65 &0.61      \\
	PPS5$^m$     & 0.72  &0.67  & 1.19    \\
        CT7$^n$       & \underline{2.99}   &  2.18& 1.60   \\
        S22$^o$      &0.92   &1.20  & 1.61       \\
        WATER27$^p$&7.99    &5.76& 4.72         \\

\hline
	TMAE  &4.08&3.66&3.98  \\
	\hline

\hline\hline
\end{tabular}
\begin{flushleft}
 (a) 6 atomization energies~\cite{Peverati20120476,doi:10.1021/jp035287b,haunschild2012theoretical}, (b) atomization energies of 148 molecules~\cite{doi:10.1063/1.476538}, 
 (c) electron affinity of 13 molecules~\cite{quest2014peverati}, (d) ionization potential of 13 molecules~\cite{quest2014peverati},
 (e) 8 proton affinities~\cite{quest2014peverati}, (f) 9 difficult cases~\cite{Peverati20120476,doi:10.1021/ct3002656}, (g) 7 hydrocarbon chemistry test sets~\cite{Peverati20120476,doi:10.1021/jz200616w}, (h) 30 recalculated reaction energies~\cite{C7CP04913G}, (i) hydrogen transfer barrier heights of 38 molecules~\cite{Peverati20120476,
doi:10.1021/jp801805p}, 
 (j) non-hydrogen transfer barrier heights of 38 molecules~\cite{Peverati20120476,
doi:10.1021/jp801805p}, 
 (k) 6 hydrogen bonds~\cite{Peverati20120476,doi:10.1021/ct6002719}, (l) 6 dipole interactions~\cite{Peverati20120476,doi:10.1021/ct6002719}, (m) 5 $\pi-\pi$ system 
 dissociation energies~\cite{Peverati20120476,doi:10.1021/ct6002719}, (n) 7 charge transfer complexes~\cite{Peverati20120476,doi:10.1021/ct6002719}, (o) 22 non-covalent interaction test set~\cite{jurevcka2006benchmark},
 (p) 27 water  clusters binding energies~\cite{WATER27JCTC2009,WATER27JCTC2017}, (q) 9 hydrogenic~\cite{doi:10.1063/1.2370993}, and (r) 11 non-hydrogenic molecular bonds~\cite{doi:10.1063/1.2370993}.  
 \end{flushleft}
}
\end{table}
%
To rationalize the rMGGAC functional performance for small molecules, we consider several standard 
benchmark test cases and the results are 
reported in TABLE~\ref{tabthermo}. The test set is divided into three subsets, namely main group 
thermochemistry (MGT), barrier heights (BH), 
and non-covalent interactions (NCI). Inspecting the performance of different 
methods 
for MGT test cases, one can immediately 
notice that rMGGAC improves over MGGAC for AE6, IP13 and PA8 test case. Within the 
other MGT test set, 
rMGGAC is worse than MGGAC for DC9 test set. Then, for the MGT test case, 
rMGGAC performs quite similarly to SCAN, and sliglty worse than MGGAC. 
Next, for NCI test cases,
we mention that rMGGAC is considerably better than MGGAC for CT7 and WATER27 test cases,
for which
one-electron self-interaction free correlation is important.
However, for S22 and PPS5 test cases, SCAN and MGGAC perform better than rMGGAC. This 
is because SCAN and MGGAC better account for 
some amount of intermediate and
short-range vdW interactions~\cite{sun2015strongly}. 
Finally, we mention 
that the performance of rMGGAC can be further enhanced 
for PPS5 and S22 with a long-range vdW interaction correction.

\begin{figure}
\includegraphics[width= 7 cm, height = 5 cm]{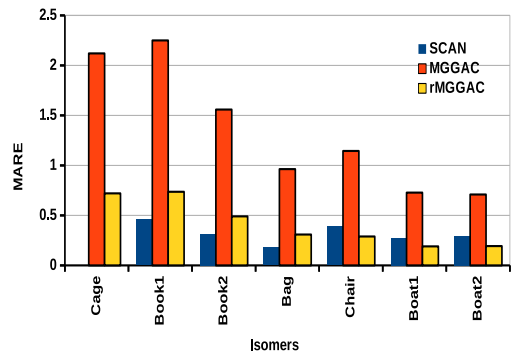}
\caption{MAREs of the different functionals for
the water hexamers with respect to the prism isomer.}
\label{fighex}
\end{figure}
%

To show the impact of the rMGGAC improvement for HB6, CT7, and DI6, in Ref.~\onlinecite{jana2020supplimentary}, 
we also calculate the relative energy differences of the water hexamers. 
The deviation of the different isomers from the prism is also shown in Fig.~\ref{fighex}. From 
Ref.~\onlinecite{jana2020supplimentary} and Fig.~\ref{fighex} it is evident that all functionals correctly 
predict the prism isomer as the most stable water hexamer, whereas the rMGGAC significantly improves over 
MGGAC for the relative energies (with respect to the prism isomer) of different isomers.

\section{SUMMARY AND CONCLUDING REMARKS}
We propose a revised MGGAC (rMGGAC) XC
functional by combining a one-electron 
self-interaction free correlation functional with the original 
MGGAC exchange. The rMGGAC correlation functional has been constructed to satisfy important exact 
conditions, being exact for one-electron systems, accurate for two-electron ground-state systems, and   
the low- and high-density limits of the uniformly scaled density. The rMGGAC correlation energy density is 
smooth and without 
spurious structure, as shown in Fig. \ref{figec}. Moreover, the rMGGAC correlation functional corrects 
the huge underestimation of the correlation energy of MGGAC for atoms and ions (see Table \ref{corr}).

The rMGGAC XC functional, as its predecessor, is an improved functional for band gap of 
strongly-bound solids, vdW solids, and 2D TMDC layers, but also improves 
considerably over MGGAC in several prospects.  
The performance of the rMGGAC demonstrates its successes in the broad direction of 
solid-state and quantum chemistry properties, which includes thermochemical, energy 
ordering for challenging polymorphs of solids and isomers of water hexamers, and  structural properties of 
solids.

Lastly, we conclude that the newly designed rMGGAC functional can be further implemented to design novel 
materials having advance 
functionalities in the field of unique device architecture. Also, the rMGGAC can offer 
opportunities for interesting 
applications of advance energy storage, catalytic problems, electronics and valleytronics.

\section*{Computational details}

We perform molecular calculations using a development version of
Q-CHEM~\cite{doi:10.1080/00268976.2014.952696} software package with def2-QZVP basis set. The XC integrals are performed   
with 99 points radial grid and 590 points angular Lebedev grid. Note that the present choice of the grid is adequate for the complete energy convergence of the non-bonded systems. 

Solid-state calculations are performed in Vienna \textit{Ab initio} Simulation Package
(VASP) ~\cite{PhysRevB.47.558,PhysRevB.54.11169,
PhysRevB.59.1758,KRESSE199615}. For all the simulations, we use $20\times 20\times
20$  $\bf{k}$ points sampling with energy cutoff $800$ eV.
For cohesive energies we consider an orthorhombic box of size  $23\times 24\times 25$
\AA$^3$ is considered. The $3^{rd}$ order Birch-Murnaghan equation of state is used to calculate the bulk moduli.
The surface energy and CO adsorption energies are calculated with $16 \times 16 \times 1$ $\Gamma$-centered $k$ 
points with an energy cutoff of $700$ eV and a vacuum level of $20$ \AA~. For CO adsorption, we relax the top two layers.

\section*{Supporting information}

The supporting information is attached with the manuscript and available online. Additional details
are available to the corresponding author upon reasonable request.

\section*{Acknowledgments}

S.J. is grateful to the NISER for partial financial
support. S.K.B acknowledges NISER for financial support. 
Calculations are performed in the 
KALINGA and NISERDFT High Performance Computing Facility, NISER.
Part of the simulations and/or computations are also supported by the 
SAMKHYA: High Performance Computing Facility provided by Institute of 
Physics (IOP), Bhubaneswar. S.J. and P.S. grateful to Prof. Suresh Kumar Patra for the computational facility at IOP, Bhubaneswar. 
S.J. and P.S. would like to thank Q-Chem, Inc. and developers for providing the source code. S.\'S. is grateful to the National Science Centre, 
Poland for the financial support under Grant No. 2020/37/B/ST4/02713

\appendix

\section{Low density and high density limit of rMGGAC correlation functional}

$(a)$ In low density limit ($r_s\to \infty$), the rMGGAC exchange-correlation (xc) energy functional should independent of 
$\zeta$ for $0\leq|\zeta|< 0.7$, a constraint important for atomization energies. In the low-density 
limit the LDA correlation energies behaves as,
\\
\begin{eqnarray}
\epsilon_c^{LDA0}(r_s)&\to^{r_s\to \infty}& -\frac{c_{1}}{r_s}+\frac{c_{2}}{r_s^{3/2}}+O(r_s^{-2})\\
\epsilon_c^{LSDA1}(r_s,\zeta) &\to^{r_s\to \infty}& -\frac{d_0(\zeta)}{r_s}+\frac{d_1(\zeta)}{r_s^{3/2}}+O(r_s^{-2})~,
\end{eqnarray}
where the coefficients related to $\epsilon_c^{LDA0}(r_s\to\infty)$ becomes $c_1=\frac{b_{1c}}{b_{3c}}$ and $c_2=\frac{b_{1c}b_{2c}}{b_{3c}^2}$.
Similarly, we find that $d_0(\zeta)$ and $d_1(\zeta)$ becomes~\cite{perdew1992accurate},
\begin{eqnarray}
 d_0(\zeta) &=& d_{xc}(\zeta)-\frac{3}{4\pi}(\frac{9\pi}{4})^{1/3}d_x(\zeta)~,\\
d_1(\zeta) &=& 1.5
 \end{eqnarray}
where
\begin{eqnarray}
d_{xc}&=&0.4582d_x+0.4335-0.1310f(\zeta)+0.0262f(\zeta)\zeta^4\nonumber\\
\end{eqnarray}
and
\begin{eqnarray}
d_x(\zeta)&=&[(1+\zeta)^{4/3}+(1-\zeta)^{4/3}]/2
\end{eqnarray}
with $f(\zeta)=\frac{d_x-1}{2^{1/3}-1}$. Note that $d_1(\zeta)$ is obtained from the spin-independent 
coefficient from the PC model~\cite{seidl2000density}.

Finally, in the low density limit the rMGGAC correlation energy becomes,
\begin{eqnarray}
W_{\infty}^{rMGGAC}&=&E_x^{MGGAC}[n_{\uparrow},n_{\downarrow}]-\int~d^3r~n({\bf{r}})[(\frac{c_1}{r_s}\nonumber\\
& &\frac{1}{[1+4\chi_{\infty}(\zeta=0)s^2]^{1/4}})G_c(\zeta)f_1(\alpha,p)\nonumber\\
&+&(\frac{d_0(\zeta)}{r_s}\frac{1}{[1+4B_1(r_s,\zeta)t^2]^{1/4}})f_2(\alpha,p)]~,\nonumber\\ 
\end{eqnarray}
with $\chi_{\infty}(\zeta=0)=0.128026$, $B_1(r_s,\zeta)=\frac{\beta_0}{d_0(\zeta)}\phi(\zeta)^3r_s$, and 
$t=(3\pi^2/16)^{1/3}s/(\phi(\zeta) r_s^{1/2})$. For details of the calculations and notations see 
main text and refs.~\cite{sun2015strongly,seidl2000density}.
\vspace{1 cm}

$(b)$ Under Levy's uniform scaling~\cite{levy1985hellmann} to the high-density limit i.e., $n_\lambda({\bf{r}})=\lambda^3 n(\lambda{\bf{r}})$ with $\lambda\to \infty$ [$\lambda\to \infty$ and $r_s\to 0$]
the $\epsilon_c^{rMGGAC}$ tends to 
\begin{equation}
\epsilon_c^{rMGGAC}\to\epsilon_c^{rMGGAC-GL2} =\epsilon_c^{0GL2}f_1(\alpha,p)+\epsilon_c^{1GL2}f_2(\alpha,p)~,
\end{equation}
where $\epsilon_c^{0GL2}$ and $\epsilon_c^{1GL2}$ are the G\"{o}rling-Levy (GL2)~\cite{gorling1994exact,gorling1993correlation,gorling1992requirements} limit of the correlation energy densities. Note that $f_1(\alpha,p)$ 
and $f_2(\alpha,p)$ remain invariant under the uniform scaling. Performing the Taylor expansion in the limit $\lambda\to \infty$ the 
resultant $\epsilon_c^{0GL2}$ and $\epsilon_c^{1GL2}$ are obtain as,

\begin{equation}
    \epsilon_c^{0GL2} = b_{1c}G_c(\zeta)\ln\{1-g_{\infty}(\zeta=0,s)(\exp(1)-1)/\exp(1)\},
\end{equation}

and 

\begin{equation}
    \epsilon_c^{1GL2} = \gamma\phi^3\ln[1-\frac{1}{(1+4\chi\frac{s^2}{\phi^2})^{1/4}}],
\end{equation}
with $\chi=(\beta_0/\gamma) c^2 e^{-\omega/\gamma} \approx 0.72161$, $\beta_0=0.066725$, $\gamma=0.031091$, $c=1.2277$, 
and $\omega=0.046644$. See ref.~\cite{perdew1996generalized} for details of the parameters.

\twocolumngrid
\bibliography{reference.bib}
\bibliographystyle{apsrev4-1}

\end{document}